\documentclass[12pt,a4paper]{iopart}
\usepackage[dvips]{graphicx} 
\usepackage{bm}
\begin{document}
\title{Effective field theory of the classical two-dimensional
 plasma} 

\author{M A Valle Basagoiti}
 
\address{Departamento de F\'\i sica
Te\'orica, Universidad del Pa\'\i s Vasco, Apartado 644, E-48080
Bilbao, Spain} 

\ead{wtpvabam@lg.ehu.es}

\date{\today}

\begin{abstract}
    Using techniques 
    of effective field theory, we consider the thermodynamical properties of 
    a dilute two-dimensional plasma interacting via a $1/r$
    potential.  The first one-loop correction to the partition function is
    already logarithmically divergent in the effective theory.  
    The finite part
    of the one-loop induced couplings is then explicitly evaluated by matching the
    density-density correlator in the effective theory and in the full
    quantum theory. This task is performed from the formulation of the Coulomb 
    problem in momentum space   
    by projecting the two-dimensional 
    momentum space onto the surface of a three-dimensional sphere.
    We also report some higher order results 
    that, in the case of the one-component  plasma immersed in 
    a uniform neutralizing background, are resummed to obtain the 
    complete leading-log  partition function. 
\end{abstract}

\submitto{\JPA}

\pacs{11.10.Kk, 52.27.-h, 3.65.-w}

\maketitle

\section{\label{Intro} Introduction}

An application of effective field theory methods, which have become commonplace in 
high-energy physics and nuclear physics, has been achieved very recently by 
Brown and Yaffe~\cite{Brown} to 
analyze some general equilibrium properties of non-relativistic classical plasmas. 
In an exhaustive work, these authors have systematically derived  a couple of 
higher-order results concerning with the thermodynamics and 
have computed some correlation functions
of the three-dimensional classical plasma. Although of some of those results were known, 
the computation is streamlined by the organization inherent in the technique of 
effective interactions. 

The two-dimensional electron gas with $1/r$ interactions is both a fundamental model 
in many-body physics and a system of current experimental interest. 
Electrons trapped on the liquid-helium surface or electrons confined in 
the vicinity of a junction between a semiconductor and insulators or 
between layers of different semiconductors are examples of two-dimensional  
electron systems \cite{Ando}. 
At finite temperature, the  parameter $n \lambda^2$
measures the dilution of the system and   
the dimensionless coupling parameter defined by $g=2\pi n \beta^2 e^4$ characterizes 
the strength of Coulomb interactions, where $\lambda$ denotes the thermal wavelength and 
$e$ is the effective electronic charge in unrationalized units. For electrons on a 
helium film the electron density can be varied in the range from $10^5$ cm$^{-2}$ to 
$10^9$ cm$^{-2}$. For these low densities, it appears that the electrons form 
classical two-dimensional systems even at temperatures of a few millidegrees Kelvin. 
For $T \sim 1$ K and $10^5 < n < 10^6$ cm$^{-2}$, the range of the plasma parameter  
is $1.7 < g < 17$, not very far from the domain of weak coupling. 
  
In this paper, we largely pursue the methods advocated by Brown and Yaffe, and
consider a classical two-dimensional multi-component plasma interacting 
via the Coulomb potential.  
From the appropriate scalar field theory, 
we compute the partition function and the correlation functions to 
first nontrivial order requiring renormalization. 
The analytical results that we have derived are valid in the low density and 
weak coupling domain where $n \lambda^2 \ll 1$ and $g \ll 1$. 
The essential ingredients in the calculation are the matching coefficients or 
induced couplings, whose finite parts are determined from a comparison of the short-distance 
behaviours in the effective  theory and in the full quantum theory. 
The main result of this paper is precisely the computation of the matching coefficients, 
exploiting some properties in the momentum space of the Coulomb Green's function 
for the quantum problem.     

There are old 
analytical and numerical results on the two-dimensional one-component plasma in the classical regime 
\cite{Fetter,Chalupa,Totsuji,Chester}. 
Chalupa \cite{Chalupa} reports an equation of state given by 
\begin{equation}
\frac{\beta p}{n} - 1 = \frac{g}{4} \ln g + 
  \frac{g}{4}\,\left(2\gamma_{\mathrm{E}} - 1 + \ln 2 \right)+ O(g^2 \ln^2 g,g^2 \ln g, g^2), 
\end{equation}
and Totsuji \cite{Totsuji} gives the ratio of the correlation energy density to the kinetic energy
as
\begin{equation}
\frac{\beta E_{\mathrm{c}}}{n}=\frac{\beta u}{n}-1 = \frac{g}{2} \ln g + 
  \frac{g}{2}\,\left(2\gamma_{\mathrm{E}} - 1 + \ln 2 \right)+ \ldots. 
\end{equation}
These expressions  agree precisely with our results (\ref{eq:pres}) and 
(\ref{eq:interna}) for the pressure and the internal energy  when 
the ratio of the Coulomb energy of a 
pair of charged particles separated by a thermal length to the 
thermal energy is very large, $\eta = \beta e^2/\lambda \gg 1$. 
However, as we will show below, our one-loop results do not imply restriction to 
this ratio. They are valid to order $g$ for arbitrary $\eta$.  
The terms of order $g^2$ retained by Chalupa and Totsuji   
are not reliable because they 
must be computed by matching with unknown results of a three-body calculation in the quantum theory. 
However, we have been able to compute all the $g^n (\ln g)^n$ leading logarithmic terms 
of the partition function in (\ref{eq:leadlog}).

The plan of this paper is arranged as follows. In section~\ref{sec:oneloop} we briefly review the
structure of the effective theory, the partition function to one-loop order and write down the 
form of the induced couplings. In Section~\ref{sec:results}, 
with the matching procedure completed, 
we compute at one-loop level the equation of state, the internal energy and the number 
density correlators. The section~\ref{sec:higher} is devoted to report some 
higher order results concerning the leading logarithmic contribution to the partition function. 
A short conclusion is presented in section~\ref{sec:conclu}. 
Details of the derivation of the basic formulae for the matching procedure are given in 
the appendix. 

\section{\label{sec:oneloop} One-loop divergences in the effective theory} 

We consider a plasma of different species of charged
particles interacting through the Coulomb interaction. The charge,
mass,\ldots of the species $a$ are denoted by $e_{a}$, $m_{a}$,\ldots. 
These are the conventions of \cite{Brown} which we shall closely follow. 
In the classical limit, the free-particle distribution function 
reduces to a  gaussian function $f_a^0(\bm{p})=e^{\beta \mu_a} e^{-\beta p^2/2m_a}$, 
and the free-particle density in two dimensions is 
$n_a^0= g_a \lambda_a^{-2} e^{\beta \mu_a}$, where 
\begin{equation}
\lambda_a= \left(\frac{2\pi \beta \hbar^2}{m_a} \right)^{1/2}
\end{equation}
denotes the thermal wavelength of the species $a$ and $g_a, \mu_a$ are the 
corresponding degeneracy and chemical potential. 
We shall assume that the system is electrically neutral. 

The Fourier transform of the $1/r$ potential in two
dimensions is
\begin{equation}
 \int d^2\bm{r} \frac{e^{-i \bm{k}\cdot\bm{r}} }{r}=\frac{2\pi}{k} \,,
\end{equation}
which gives rise to 
the interaction potential in $\nu$ spatial dimensions 
\begin{equation}
 V_{\nu}(\bm{r}-\bm{r}') = \int \frac{d^\nu\bm{k}}{(2\pi)^\nu}
 \frac{e^{i \bm{k}\cdot(\bm{r}-\bm{r}')} 2\pi}{k} = 
 \Gamma\left(\frac{\nu-1}{2}\right)\pi^{(1-\nu)/2}|\bm{r}-\bm{r}'|^{1-\nu}\,.
\end{equation}
As usual, in the absence of any scale, 
the analytically continued coincidence limit $V_{\nu}(\mathbf{0})$ will be taken as zero.  
Using the fact that $\sqrt{-\nabla^2}/(2\pi)$ is the inverse operator of 
$1/r$ in two dimensions, 
the grand canonical partition function in the classical limit can be 
turned via a Hubbard-Stratonovich 
transformation into the functional integral  
\begin{equation}\label{eq:funct}
\mathcal{Z}(\mu)=\mathrm{Det}^{1/2}\left[ \frac{\beta \sqrt{-\nabla^2}}{2\pi}\right]
\int \mathcal{D}\phi(\bm{r}) \exp \left(-S_{\mathrm{cl}}[\phi;\mu] \right), 
\end{equation}
with the action functional of the electrostatic potential $\phi$ 
defined by 
\begin{equation}\label{eq:action}
    S_{\mathrm{cl}}[\phi;\mu] = \int d^2 \bm{r}
\left\{ \frac{\beta}{2} \phi(\bm{r})
\frac{\sqrt{-\nabla^2}}{2\pi}\phi(\bm{r})- \sum_{a} n_{a}^0(\bm{r})
e^{i\beta e_{a} \phi(\bm{r})}\right\}.
\end{equation}
where $n_{a}^0(\bm{r}) = g_a \lambda_a^{-2} e^{\beta \mu(\bm{r})}$. 
The corresponding field equation is the analog of the   
Debye-H\"uckel result in two dimensions.   
Here, it proves convenient to assume that the chemical potentials entering into $n_a^0$ 
can depend upon $\bm{r}$. Thus, the functional derivation with respect to $\mu_a(\bm{r})$ 
will produce the correlation function between densities 
\begin{equation}\label{eq:correl}
K_{ab}(\bm{r}-\bm{r}') = 
\frac{\delta^2\ln\mathcal{Z}(\mu)}{\delta\beta \mu_a(\bm{r})\delta\beta \mu_b(\bm{r}')}.
\end{equation} 
Since the total charge neutrality in terms of the free-particle densities is 
\begin{equation}
\sum_a e_a n_a^0 = 0,  
\end{equation}     
the $\phi=0$ configuration is a solution of the classical field equation and,
consequently, the functional integral can be perturbatively computed from a saddle point expansion 
around this trivial solution. 
The quadratic part of the action,   
\begin{equation}\label{eq:quadratic}
S_{0}[\phi;\mu] = \int d^2 \bm{r} \left\{- \sum_{a=1}^A n_{a}^0
+\frac{\beta}{2} \phi(\bm{r}) \left[\frac{\sqrt{-\nabla^2}}{2
\pi} + \frac{\kappa_{0}}{2\pi}\right] \phi(\bm{r}) \right\} ,
\end{equation}
includes the inverse Debye length, $\kappa_0\equiv 2 \pi \beta \sum_a e_a^2 n_a^0$,  
and the remaining part giving rise to the perturbative expansion is 
\begin{equation}\label{eq:intera}
\Delta S[\phi;\mu] = -\int d^2 \bm{r} \sum_{a=1}^A n_{a}^0 \left\{
e^{i\beta e_{a} \phi(\bm{r})}-1 + 
\frac{1}{2} \beta^2 e_{a}^2 \phi(\bm{r})^2  \right\}.
\end{equation}

This effective field theory can   
be systematically obtained from the quantum statistical mechanics of the plasma 
by integration of the quantum fields with momentum scales of order $\lambda^{-1}$ 
or larger~\cite{Brown}. 
In addition to the terms already present in (\ref{eq:action}), there are
sub-leading contributions 
in a derivative expansion in
powers of $(\lambda \partial \phi)$.  The first of these comes from
the static non-interacting two-point charge density-charge density
correlation at non-vanishing $\bm{k}$ and zero frequency, evaluated
within the Maxwell-Boltzmann statistics.  This reads
\begin{equation}
\Pi(\bm{k},\omega=0) =  \frac{\kappa_0}{2\pi} 
\left(1 -\frac{\lambda^2 k^2}{24\pi} +\ldots \right),   
\end{equation} 
where the piece independent upon $\bm{k}$ produces the Debye screening
included in (\ref{eq:action}) and the next term yields an action
contribution linear in the particle densities, 
\begin{equation}\label{eq:deriv}
    -\int d^2 \bm{r} \frac{\beta
\kappa_{0}\lambda^2}{96\pi^2}\,
\nabla\phi(\bm{r})\cdot\nabla\phi(\bm{r}). 
\end{equation}
This action must be made consistent with the invariance of the original
theory under the simultaneous constant shift $\phi\rightarrow \phi- i
c$ and $\mu_{a}\rightarrow \mu_{a}-e_{a} c$.  Consequently, the
corresponding derivative interaction, linear in the particle densities must
have the form
\begin{equation}\label{eq:indu1}
 S_{\mathrm{ind}}^{(1)}[\phi;\mu]=\sum_{a}\int d^2 \bm{r}\, 
 \frac{\beta^2\lambda_{a}^2}{48\pi} \left[ 
 \nabla\left(\mu_{a}(\bm{r}) + i e_{a}\phi(\bm{r})\right) \right]^2
 n_{a}^0(\bm{r}) e^{i\beta e_{a} \phi(\bm{r})}, 
\end{equation}
where we have used the form of the Debye wave number.

In the following, 
we safely use the effective theory of equation (\ref{eq:action}) to describe distance
scales of order $\lambda$ and larger.  Although this classical regime is
independent of the quantum statistics, the logarithmic ultraviolet
divergences in the effective theory are
related to quantum properties of the full theory. These divergences of the 
form $\ln(\kappa_0 /\mu)$, where $\mu$ is a
momentum scale introduced by dimensional regularization, 
must match onto counterterms proportional to $\ln(\mu \lambda)$
arising from the action with the
induced couplings required to cancel the $\mu$ dependence.
The precise value of the finite part of the induced couplings 
is determined by comparison with the
result of the computation of some quantity in the full quantum theory. 
The main result of this work is justly the explicit evaluation of the
one-loop induced couplings.
 
The dimensionless loop expansion parameter is the ratio of the Coulomb 
energy for two particles 
separated by a Debye length to their average kinetic energy in the plasma. 
This plasma parameter is  $g=\beta e^2 \kappa_0$. 
Apart from the Debye length,
there is another relevant length scale, not directly entering into the effective theory, 
but playing an important role in 
the matching procedure.  
This is the Bohr radius $(e^2 m_{ab})^{-1}$ or, equivalently, 
the binding energy $e^4 m_{ab}/2$ of two particles in the plasma with 
reduced mass $m_{ab}$ and equal and opposite charge.

The Green's function $G_\nu(\bm{r}-\bm{r}')$ in $\nu$ dimensions, 
including the effects of static screening, is given in terms of Struve functions
$\mathbf{H}$ and 
Bessel functions of second kind $Y$ by 
\begin{equation}
G_\nu(\bm{r}-\bm{r}')=\int \frac{d^\nu\bm{k}}{(2\pi)^\nu}
  \frac{e^{i \bm{k}\cdot(\bm{r}-\bm{r}')} 2\pi}{k+\kappa_0} = 
 V_\nu(\bm{r}-\bm{r}') + \Delta V_\nu(|\bm{r}-\bm{r}'|),
\end{equation}
with 
\begin{equation}
\Delta V_\nu(r) = \frac{2^{-\nu/2} \pi^{2-\nu/2} \kappa_0^{\nu/2} r^{1-\nu/2}}{\cos(\pi\nu/2)}
\left[\mathbf{H}_{1-\nu/2}(\kappa_0 r) - Y_{1-\nu/2}(\kappa_0 r) \right],   
\end{equation}
and the coincidence limit is 
\begin{equation}\label{eq:G0}
 G_\nu(\mathbf{0}) = 
 \frac{2^{2-\nu} \pi^{2-\nu/2} \kappa_0^{\nu-1}}{\Gamma(\nu/2) \sin\pi \nu}  .
\end{equation}
With these results in hand, the one-loop contribution to $\ln \mathcal{Z}/\mathcal{A}$ for 
the effective theory 
is given by  
\begin{eqnarray}
\frac{\ln \mathcal{Z}}{\mathcal{A}} &=& \sum_a n_a^0 - 
\frac{1}{2 \mathcal{A}}\ln\mathrm{Det}
\left[1+\frac{1}{\sqrt{-\nabla^2}} \kappa_{0}\right] \nonumber \\ 
&=& \sum_a n_a^0 - \frac{\kappa_0}{4\pi \nu}\, G_\nu(\mathbf{0}) , 
\end{eqnarray}
where the determinant has been evaluated by integration of the formula
\begin{equation}
\delta \ln \mathrm{Det} \left[1+\frac{1}{\sqrt{-\nabla^2}} \kappa_{0} \right] = 
\int d^\nu \bm{r} G_{\nu}(\mathbf{0}) \frac{\delta\kappa_{0}}{2\pi}. 
\end{equation}
This contribution is singular when 
$\nu\rightarrow 2$ and, already  to this order, renormalization is required. To 
determine the form of counterterms, we note that the
pole term is proportional to $\kappa_0^2$. Thus, at this order, 
the induced couplings 
must be quadratic in the particle densities.  The counterterm action, to be handled at tree level to 
cancel the one-loop divergences,  
must have the 
form~\cite{Brown}
\begin{equation}\label{eq:indu2}
 S_{\mathrm{ind}}^{(2)}[\phi;\mu]=\sum_{a, b} \int d^\nu {\mathbf r}\,
 \beta^2 g_{a b}^0 n_{a}^0 n_{b}^0 e^{i \beta e_{a}
 \phi(\mathbf{r})} e^{i \beta e_{b} \phi(\mathbf{r})} ,
\end{equation}
where the pole term of $g_{ab}^0$
\begin{equation}\label{eq:gab0}
g_{ab}^0 = \mu^{\nu-2}\left[\frac{\pi e_a^2 e_b^2 }{2}\frac{1}{2-\nu} + g_{ab}(\mu) \right], 
\end{equation}
cancels the $\nu\rightarrow 2$ pole of $\ln \mathcal{Z}$ coming from $G_2(\mathbf{0})$. 
The momentum scale $\mu$ is required in order for the finite part of the coupling $g_{ab}(\mu)$
preserve its dimensions at $\nu=2$ when $\nu$ is arbitrary. To this order, 
the contribution of the induced couplings only involves the 
$\phi=0$ part of $S_{\mathrm{ind}}$ but at higher orders,
when it is 
neccesary to handle the induced interaction 
beyond the tree level, the full exponential dependence upon $\phi$ will be  
crucial. 
 
The renormalization group equation 
for $g_{ab}(\mu)$,
\begin{equation}
\mu \frac{d}{d\mu} g_{ab}(\mu) = \frac{\pi e_a^2 e_b^2}{2},
\end{equation}
guarantees that the bare induced couplings $g_{ab}^0$ do not depend on $\mu$.
By integration, one obtain 
\begin{equation}\label{eq:muab}
    g_{ab}(\mu) = \frac{\pi e_a^2 e_b^2}{4}
\ln\left(\frac{\mu^2}{\mu_{ab}^2}\right),
\end{equation} 
where the integration constant $\mu_{ab}$ cannot be determined within the effective theory. 
The  simplest way to determine the short-distance parameter $\mu_{ab}$ is 
to match the quadratic part in the particle densities of the one-loop density-density correlator 
$\widetilde{K}_{ab}(\bm{k})$ in the effective theory and in the full quantum theory~\cite{Brown}.
In this comparison of the short-distance behaviours, 
the Debye screening plays no role~\cite{Nieto} provided that the correlators 
are evaluated at non-zero wave number $\bm{k}$. 

\begin{figure}
    \includegraphics{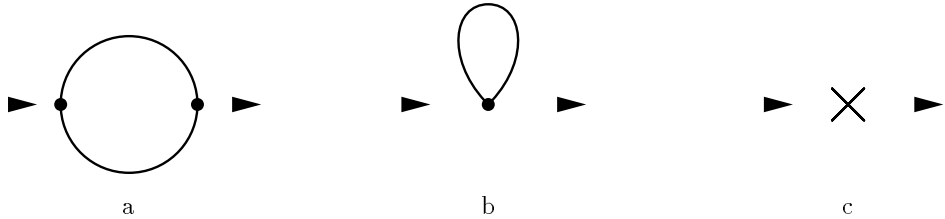}
    \caption{\label{fig:oneloop} One-loop diagrams contributing to 
    the irreducible part $\widetilde{C}_{ab}(\bm{k})$ of the density-density
    correlator. The diagram c represents the contribution of induced couplings.}
\end{figure}

As discussed by Brown and Yaffe \cite{Brown}, the Fourier transform of the complete 
density-density correlator has the general form 
\begin{equation}
\widetilde{K}_{ab}(\bm{k})=\widetilde{C}_{ab}(\bm{k}) - 
\frac{\left(\beta \sum_c e_c \widetilde{C}_{ca}(\bm{k}) \right) \, 2\pi 
\left(\beta\sum_c e_c \widetilde{C}_{cb}(\bm{k})\right)}
{\beta(|\bm{k}|+ 2\pi \beta \sum_{m n}e_m e_n \widetilde{C}_{mn}(\bm{k}))}. 
\end{equation}
Here, $\widetilde{C}_{ab}(\bm{k})$ denotes the Fourier transform of 
\begin{equation}
C_{ab}(\bm{r}-\bm{r}')=-\left.
\frac{\delta^2 \Gamma[\overline{\phi}; \mu]}{\delta\beta \mu_a(\bm{r}) \delta\beta \mu_b(\bm{r}')}\right|_
{\overline{\phi}=0} \,,
\end{equation}
where $\Gamma[\overline{\phi}; \mu]$ is the effective action of the theory. At tree level, 
$\Gamma[\overline{\phi}; \mu] =S_{\mathrm{cl}}[\overline{\phi};\mu]$ and, consequently, 
$\widetilde{C}_{ab}^{\mathrm{tree}}(\bm{k})= \delta_{ab} n_a^0$ (no sum over $a$). 
Noting that $S_{\mathrm{ind}}^{(1)}[\phi;\mu]+S_{\mathrm{ind}}^{(2)}[\phi;\mu]$ 
in equations (\ref{eq:indu1}) and (\ref{eq:indu2}) must be handled 
at tree level,   
the contribution to  
$\widetilde{C}_{ab}(\bm{k})$ up to one-loop order is given by
\begin{eqnarray}\label{eq:irreducible}
\widetilde{C}_{ab}^{(1)}(\bm{k})&=& \delta_{ab} n_b^0 +
\beta^2 n_a^0 n_b^0 \left[\frac{e_a^2 e_b^2}{2} D_\nu^{(2)}(\bm{k})-2 g_{ab}^0\right] -
\frac{1}{2}\beta \delta_{ab} e_b^2 n_b^0 G_{\nu}(\mathbf{0}) \nonumber \\ 
&&- 2 \beta^2 \delta_{ab}\sum_c g_{bc}^0 n_b^0 n_c^0 
   -\frac{\lambda_a^2 k^2}{24\pi}\,\delta_{ab} n_b^0 ,
\end{eqnarray}
where 
\begin{equation}
D_\nu^{(2)}(\bm{k})\equiv \int d^\nu {\bm{r}} e^{-i\bm{k}\cdot \bm{r}} G_{\nu}(\bm{r})^2 . 
\end{equation}
The function $D_\nu^{(2)}(\bm{k})$ comes from the one-loop diagram a in figure~\ref{fig:oneloop} 
and $G_{\nu}(\mathbf{0})$ is the tadpole. 
Since the last term of equation~(\ref{eq:irreducible}) 
yields a contribution to the $n_a^0 n_b^0$ part of $\widetilde{K}_{ab}(\bm{k})$ 
which vanishes linearly when $k\rightarrow 0$, this term 
can be ignored in the matching procedure at small momentum $\bm{k}$. 
Therefore, ignoring Debye screening,  
\begin{equation}
\fl
D_\nu^{(2)}(\bm{k}) \rightarrow C_\nu^{(2)}(\bm{k}) 
\equiv \int d^\nu {\bm{r}} e^{-i\bm{k}\cdot \bm{r}} V_{\nu}(\bm{r})^2  = 
\frac{2^{2-\nu} \pi^{1-\frac{\nu}{2}}\,\Gamma\left(1-\frac{\nu}{2}\right) 
\Gamma\left(\frac{1-\nu}{2}\right)^2}{\Gamma(\nu-1)} k^{\nu-2} . 
\end{equation} 
This replacement leads to the piece of the correlator to be compared to the corresponding result in 
the full quantum theory
\begin{eqnarray}\label{eq:quad}
\fl
\widetilde{K}_{ab}^{\mathrm{quad}}(\bm{k})=
\beta^2 n_a^0 n_b^0 \left[\frac{e_a^2 e_b^2}{2} C_\nu^{(2)}(\bm{k})-2 g_{ab}^0\right]-
\frac{2\pi \beta n_a^0 n_b^0 e_a e_b}{k} \nonumber \\ 
\lo=-\beta^2 n_a^0 n_b^0 \left\{ \frac{\pi e_a^2 e_b^2}{2}\left[
\gamma_{\mathrm{E}} + \ln\left(\frac{k^2}{64 \pi \mu^2}
\right)\right]+2 g_{ab}(\mu)\right\}- \frac{2\pi \beta n_a^0 n_b^0 e_a e_b}{k}  .
\end{eqnarray}


\section{\label{sec:results} Matching and one-loop complete results}

In order to perform the matching, we consider the $\bm{k}\rightarrow 0$ behaviour of the
second-order contribution in the fugacity expansion to the quantum
mechanical density-density correlator $\mathcal{K}_{ab}(\bm{k})$. 
The details of the computation are given in the appendix. 
This contribution to the full correlator is 
\begin{equation}\label{eq:quantum}
    \mathcal{K}_{ab}^{(2)}(\bm{k}) = n_{a}^0 n_{b}^0
\lambda_{ab}^2 \left[F_{+}(\bm{k}) \pm (\delta_{ab}/g_{a})
F_{-}(\bm{k}) \right],
\end{equation}
where the $\pm$ sign accounts for Bose ($+$) or Fermi ($-$) statistics,
$g_{a}$ denotes the spin degeneracy of the species $a$, and
$\lambda_{ab}$ is the thermal wavelength for the reduced mass
$m_{ab}=m_{a} m_{b}/(m_{a}+m_{b})$.  The required terms are (Fourier
transformed) matrix elements of the heat kernel constructed in terms
of the Hamiltonian $H$ for relative motion, and read
\begin{equation}
 F_{\pm}(\bm{k})= \int d^2 \bm{r} e^{-i \bm{k}\cdot\bm{r}} 
 \langle\bm{r}|e^{-\beta H_{ab}} |\pm\bm{r}\rangle ,
\end{equation} 
with
\begin{equation}
 H_{ab} = \frac{\bm{p}^2}{2m_{ab}} + \frac{e_{a}e_{b}}{r} .
\end{equation} 

With the results (\ref{eq:fplus}) and (\ref{eq:fminus}) in hand, the
comparison of (\ref{eq:quantum}) with (\ref{eq:quad}) enables
to determine the integration constant $\mu_{ab}$ in the running
coupling
\begin{equation}
 g_{ab}(\mu) = \frac{\pi e_{a}^2 e_{b}^2}{4}\left[ \ln(\mu^2
 \lambda_{ab}^2) + f(\eta_{ab}) \pm (\delta_{ab}/g_{a})
 \widetilde{f}(\eta_{aa})\right], 
 \end{equation}
 in terms of the quantum parameter $\eta_{ab}=\beta e_{a} e_{b}/\lambda_{ab}$. 
 Thereby, 
 the  complete contribution to $\ln\mathcal{Z}$ up to
 one-loop order in terms of the fugacities is
 \begin{equation}
\frac{\ln \mathcal{Z}(\mu)}{\mathcal{A}}= \sum_a n_a^0 - 
\frac{\pi}{4}\sum_{a,b} n_a^0 n_b^0 \left(\beta e_{a}e_{b} \right)^2
\left[ \ln\left(\frac{\lambda_{ab}^2\kappa_0^2}{4\pi}\right) -1 +
\gamma_{\mathrm{E}}+ \Gamma_{ab}\right] ,
\end{equation}
where
\begin{equation}\label{eq:Gammaab}
 \Gamma_{ab}=f(\eta_{ab}) \pm (\delta_{ab}/g_{a})
\widetilde{f}(\eta_{aa}) .
\end{equation}

By inversion of the formula 
\begin{equation}
\overline{n}_a = \frac{1}{\beta}
\frac{\partial}{\partial \mu_a}\frac{\ln \mathcal{Z}(\mu)}{\mathcal{A}},
\end{equation}
the elimination of the chemical potentials in favour of the 
mean number densities  $\overline{n}_a$ can be made perturbatively, 
and at one-loop order we obtain
\begin{equation}\label{eq:mean}
n_a^0 = \overline{n}_a \left(1 +  
\frac{\pi}{2}\sum_{b} \overline{n}_b \left(\beta e_{a} e_{b}\right)^2
\left[\ln\left(\frac{\lambda_{ab}^2\overline{\kappa}_0^2}{4\pi}\right) 
+\gamma_{\mathrm{E}}+ \Gamma_{ab}\right]\right) ,
\end{equation}
with $\overline{\kappa}_0 = 2\pi \beta \sum_a e_a^2 \overline{n}_a$. 
The corresponding replacement in the expression for the 
pressure $p=\ln \mathcal{Z}/\beta$ gives 
the equation of state
\begin{equation}\label{eq:pres} 
\beta p = \sum_a \overline{n}_a+
\frac{\pi}{4}\sum_{a,b} \overline{n}_a \overline{n}_b \left(\beta e_{a}e_{b} \right)^2
\left[\ln\left(\frac{\lambda_{ab}^2\overline{\kappa}_0^2}{4\pi}\right)
+1 +\gamma_{\mathrm{E}}+ \Gamma_{ab}\right].  
\end{equation}
The internal energy $u$ density is given by
\begin{equation}
u= -\frac{\partial}{\partial\beta}
\frac{\ln \mathcal{Z}(\mu)}{\mathcal{A}}+
\sum_a  \mu_a \overline{n}_a ,
\end{equation}
which yields
\begin{equation}\label{eq:interna}
\fl
\beta u = \sum_a \overline{n}_a +
\frac{\pi}{2}\sum_{a,b} \overline{n}_a \overline{n}_b \left(\beta e_{a}e_{b} \right)^2
\left[\ln\left(\frac{\lambda_{ab}^2\overline{\kappa}_0^2}{4\pi}\right)
+\frac{1}{2} +\gamma_{\mathrm{E}}+ \Gamma_{ab} + 
\frac{1}{4}\, \eta_{a b}\Gamma_{ab}' \right],   
\end{equation}
where 
\begin{equation}
\Gamma_{ab}'=f'(\eta_{ab}) \pm (\delta_{ab}/g_{a})
\widetilde{f}'(\eta_{aa}) .
\end{equation}

In the case of a plasma consisting of a single species immersed in
a uniform neutralizing background, the terms of order $\epsilon^2$ and 
$\epsilon^2 \ln \epsilon$ which appear in the  equation of state reported by 
Chalupa~\cite{Chalupa} 
agree precisely with our result~(\ref{eq:pres}) 
when the strong repulsion result~(\ref{eq:strong}) is inserted\footnote{
The parameter $\epsilon^2$ in \cite{Chalupa} is $4 \pi \overline{n} e^4\beta^2= 
2 \overline{\kappa}_0 \lambda_{aa} \eta_{aa}$, and the 
parameter $\epsilon$ in  
\cite{Totsuji} is $2 \pi \overline{n} e^4 \beta^2$.}
into (\ref{eq:Gammaab}), neglecting 
the term of order $\eta_{aa}^{-2}$ and the exchange term~(\ref{eq:stroex}). 
The ratio of the correlation energy density to the kinetic energy 
density $\beta E_\mathrm{c}/n$ reported by Totsuji~\cite{Totsuji} also 
agrees with the term of order $e^4$ in (\ref{eq:interna}) 
when $\eta_{aa} \gg 1$ .  

\subsection*{Number density correlators}
The insertion of our findings into  (\ref{eq:irreducible}) yields 
the irreducible part of the number density correlator function to one-loop order, 
in terms of bare densities
\begin{equation}
\widetilde{C}_{ab}(\bm{k}) = \frac{1}{2} \left(\beta e_a^2 n_a^0\right)
\widetilde{F}_{ab}^0(\bm{k})\left(\beta e_b^2  n_b^0\right)+
\delta_{ab} n_a^0 \widetilde{F}_{a}^0(\bm{k}) .
\end{equation}   
Using (\ref{eq:G0}) and (\ref{eq:gab0}), the contact term reads 
\begin{equation}
\widetilde{F}_{a}^0(\bm{k})=
1 -
\frac{\pi}{2}\sum_{b} n_b^0 \left(\beta e_{a} e_{b}\right)^2
\left[\ln\left(\frac{\lambda_{ab}^2 \kappa_0^2}{4\pi}\right) 
+\gamma_{\mathrm{E}}+ \Gamma_{ab}\right]- \frac{\lambda_a^2 k^2}{24\pi},  
\end{equation}
and the counterpart $\delta_{ab} \overline{n}_a \widetilde{F}_{a}(\bm{k})$ 
in terms of the mean number densities is readily written 
with the aid of (\ref{eq:mean}), 
\begin{equation}
\widetilde{F}_{a}(\bm{k})=
1 -\frac{\lambda_a^2 k^2}{24\pi}\left(1 + 
\frac{\pi}{2}\sum_{b} \overline{n}_b \left(\beta e_{a} e_{b}\right)^2
\left[\ln\left(\frac{\lambda_{ab}^2 \overline{\kappa}_0^2}{4\pi}\right) 
+\gamma_{\mathrm{E}}+ \Gamma_{ab}\right]\right).  
\end{equation}
Since $G_2(\bm{r})^2-V_2(\bm{r})^2$ behaves as $\ln(\kappa_0 r)/r$ as $r\rightarrow 0$,  
the non-vanishing part of $D_\nu^{(2)}(\bm{k})$ when $\nu \rightarrow 2$ may be written as 
\begin{equation}
D_\nu^{(2)}(\bm{k}) = C_\nu^{(2)}(\bm{k}) + \Delta C_2^{(2)}(\bm{k}) ,
\end{equation}
where $\Delta C_2^{(2)}(\bm{k})$ is the Fourier transform of 
\begin{equation}
G_2(\bm{r})^2-V_2(\bm{r})^2 = \Delta V_2(r)\left(\Delta V_2(r) - \frac{2}{r} \right), 
\end{equation}
and the asymptotic behavior of $\Delta V_2(r)$ reads 
\begin{equation}
\Delta V_2(r) =\frac{\pi \kappa_0}{2}\left(Y_{0}(\kappa_0 r) -
\mathbf{H}_{0}(\kappa_0 r)  \right) \sim 
  -\frac{1}{r} + \frac{1}{\kappa_0^2\, r^3}+\ldots 
\end{equation} 
With this definition, we obtain
\begin{equation}
\widetilde{F}_{ab}^0(\bm{k}) =\Delta C_2^{(2)}(\bm{k})-
\pi\left[\ln\left(\frac{\lambda_{ab}^2 k^2}{64\pi}\right)
+\gamma_{\mathrm{E}}+ \Gamma_{ab}\right] . 
\end{equation}
We must emphasize that the dependence upon $\ln(k^2)$ is deceptive. 
In fact, an explicit calculation when  $\bm{k}\rightarrow 0$ shows that 
\begin{equation}
\Delta C_2^{(2)}(\bm{k}) = -2\pi \left[1 - \ln\left(\frac{k}{4\kappa_0}\right) \right] 
-\frac{\pi k^2}{\kappa_0^2} \ln \left( \frac{k}{\kappa_0}\right)+
O(k^2), 
\end{equation}
and, consequently  
\begin{equation}
\widetilde{F}_{ab}^0(\mathbf{0}) =-
\pi\left[\ln\left(\frac{\lambda_{ab}^2 \kappa_0^2}{4\pi}\right)+2
+\gamma_{\mathrm{E}}+ \Gamma_{ab}\right], 
\end{equation}
in perfect agreement with the general result 
$\widetilde{C}_{ab}(\mathbf{0})=-\partial \overline{n}_a/\partial\beta\mu_b$. 
Note that, to one-loop order,  the $n_a^0  n_b^0 $ piece of $\widetilde{C}_{ab}$ 
can be written in terms of  the mean densities by replacing $n^0 \rightarrow \overline{n}$. 

\section{\label{sec:higher} Renormalization at higher order}

The complete determination of two-loop and  higher order contributions to $\ln{\mathcal Z}$ 
requires at least to perform a difficult three-body calculation in the quantum theory.  
However, it is straighforward to 
derive renormalization group equations for the leading logarithmic pieces 
of the renormalized 
multiloop induced couplings $g_{i_1\ldots i_n}^0$ that enter into the 
counterm action through
\begin{equation}
S_{\mathrm{ind}}[\phi;\mu]=\sum_{n=2}\sum_{i_1\ldots i_n} \int d^\nu \bm{r} 
g_{i_1\ldots i_n}^0 \beta^{2(n-1)}
 n_{i_1}^0 e^{i\beta e_{i_1}\phi} \ldots n_{i_n}^0 e^{i\beta e_{i_n}\phi}. 
\end{equation}
The coupling $g_{i_1\ldots i_n}^0$ is proportional to $e^{4 (n-1)}$    
since at $\nu=2$, all factors of $n_i^0$ excepting one of them 
must be accompanied by a factor 
of $\beta^2 e^4$, in order to reproduce the right power counting. 
Thus, the induced coupling $g_{i_1\ldots i_n}^0$ first contributes 
to $\ln\mathcal{Z}/\mathcal{A}$ at order  $n-1$ loop. 
This contribution is 
\begin{equation}
I_{g_{i_1\ldots i_n}}^{(n-1)}=-\beta^{2(n-1)}\sum_{i_1\ldots i_n}  
g_{i_1\ldots i_n}^0 n_{i_1}^0  \ldots n_{i_n}^0 . 
\end{equation}
Another contribution at order  $(n-1)$-loop which arises 
from the one-loop graph with one insertion of the induced interaction 
$g_{i_1\ldots i_{n-1}}^0$ is 
\begin{eqnarray}
\fl
I_{g_{i_1\ldots i_{n-1}}}^{(n-1)}=\beta^{2(n-1)-1}\sum_{i_1\ldots i_{n-1}}  
g_{i_1\ldots i_{n-1}}^0 n_{i_1}^0  \ldots n_{i_{n-1}}^0 (e_{i_1}+\ldots+ e_{i_{n-1}})^2
\frac{1}{2} G_\nu(\bf{0})  \\
\lo{=}\beta^{2(n-1)}\pi \kappa_0^{\nu-2} \frac{1}{\nu-2}\sum_{i_1\ldots i_{n}}  
g_{i_1\ldots i_{n-1}}^0 n_{i_1}^0  \ldots n_{i_{n}}^0 
(e_{i_1}+\ldots+ e_{i_{n-1}})^2 e_{i_n}^2 \nonumber , 
\end{eqnarray}
where we have only shown the pole part of $G_\nu(\bf{0})$. 
Since the bare coupling $g_{i_1\ldots i_{n-1}}^0$ contains in general  
a pole in $\nu-2$ of order $n-2$,  
this term produces a pole contribution  of order $n-1$   
to be cancelled by $I_{g_{i_1\ldots i_n}}^{(n-1)}$. 
Moreover, there are other pole contributions to $\ln\mathcal{Z}$ generated 
by $(n-1)$-loop diagrams without induced couplings and, consequently, the cancellation 
of pole terms requires that $g_{i_1\ldots i_{n}}^0$ have the 
general form
\begin{eqnarray}
g_{i_1\ldots i_{n}}^0 &=& \pi \mu^{\nu-2}\frac{1}{\nu-2}\,
\frac{1}{n}\left(g_{i_1\ldots i_{n-1}}^0 
(e_{i_1}+\ldots+ e_{i_{n-1}})^2 e_{i_n}^2+ \textrm{permutations}\right)\nonumber \\ 
&&+\mu^{(n-1)(\nu-2)}\left[\sum_{k=1}^{n-1}\frac{R_{i_1\ldots i_{n}}^{(k)}}{(\nu-2)^k} + 
g_{i_1\ldots i_{n}}(\mu)\right], 
\end{eqnarray} 
where $R_{i_1\ldots i_{n}}^{(k)}$ comes enterely from 
contributions not included into $I_{g_{i_1\ldots i_{n-1}}}^{(n-1)}$.
Thus, the finite term at $\nu=2$ of the Laurent series for the  renormalization group condition 
$d g_{i_1\ldots i_n}^0/d\mu=0$ gives rise to 
\begin{equation}
\fl
\mu\frac{d}{d\mu}g_{i_1\ldots i_{n}}(\mu) = 
-\frac{\pi}{n}\left[g_{i_1\ldots i_{n-1}} 
(e_{i_1}+\ldots+ e_{i_{n-1}})^2 e_{i_n}^2+ \textrm{$(n-1)$ similar}\right]+
\ldots
\end{equation}
By defining the input value $g_a\equiv -1/2$, this equation recursively determines 
the leading logarithmic contribution to the induced couplings when 
$\mu$ is of order $\kappa_0$ and the integration constants are of order 
$\lambda^{-1}$. Up to three-loop order, the leading-log pieces of the partition function are
\begin{eqnarray}
\fl
\frac{\ln\mathcal{Z}(\mu)}{\mathcal{A}} =
\sum_a n_a^0 -\frac{\pi}{2}\ln (\kappa_0 \lambda) \sum_{a,b}\beta^2 
n_a^0 n_b^0  e_a^2 e_b^2   \nonumber \\
\lo 
+\frac{\pi^2}{4}\ln^2 (\kappa_0 \lambda) \sum_{a,b,c}\beta^4 
n_a^0 n_b^0 n_c^0 e_a^2 e_b^2 e_c^2 (e_a+e_b)^2  \nonumber \\
\lo 
-\frac{\pi^3}{12}\ln^3 (\kappa_0 \lambda) \sum_{a,b,c,d}\beta^6 
n_a^0 n_b^0 n_c^0 n_d^0 e_a^2 e_b^2 e_c^2 e_d^2(e_a+e_b)^2 (e_a+e_b+e_c)^2 + 
\ldots
\end{eqnarray}

In the case of one-component plasma the series of leading logs precisely agrees with the 
asymptotic expansion  for the incomplete $\Gamma(0,z^{-1})$ function, with  
$z=\pi n^0 e^4 \beta^2 \ln(\kappa_0 \lambda)$.  The result is 
\begin{eqnarray}\label{eq:leadlog}
\frac{\ln\mathcal{Z}(\mu)}{\mathcal{A}} &=& 
n^0 + \frac{n^0}{2}\sum_{j=1}^\infty (-1)^j  j! 
\left(\pi n^0 e^4 \beta^2 \ln(\kappa_0 \lambda)\right )^j  \nonumber \\  
&=& \frac{n^0}{2} \left[1 + \frac{1}{z} \exp\left(\frac{1}{z}\right) \Gamma(0,\frac{1}{z}) \right], 
\end{eqnarray} 
that produces the leading-log pressure  expressed in term of the physical number density
\begin{equation}
\fl
\frac{\beta p}{\overline{n}} -1 = \frac{1}{2}\overline{z}  
-\overline{z}^2 + \frac{5}{2}\overline{z}^3 - 11 \overline{z}^4 + 
\frac{137}{2}\overline{z}^5 - 
510 \overline{z}^6 +
4341 \overline{z}^7+O(\overline{z}^8),
\end{equation}
where $\overline{z}=\pi \overline{n} e^4 \beta^2 \ln(\overline{\kappa}_0 \lambda)$.
Finally, within this approximation the internal energy density becomes 
\begin{equation}
u = 2 p -\frac{\overline{n}}{\beta}.
\end{equation}

\section{\label{sec:conclu} Conclusion}
With Brown and Yaffe as a guide, we have shown how the technique of effective field theory can be used in order to 
compute the partition function of the classical two-dimensional plasma interacting via a $1/r$ potential. 
Renormalization in the effective theory is essential already to one-loop level, and two-body induced couplings
are required to order $e^4$. 
We have performed in detail the computation of the finite part of these couplings by 
exploiting the properties of the Coulomb Green's function in momentum space. 
We have derived explicit expressions for the one-loop equilibrium properties, 
without restriction  on the quantum parameter $\eta_{ab} = \beta e_a e_b/\lambda_{ab}$. 
The higher order induced couplings satisfy renormalization group equations which can be solved in the leading 
logarithmic approximation. In the case of one-component plasma, we have been able to derive the asymptotic series 
of leading logs for the partition function in terms of the incomplete Gamma function. 
   

\section*{Acknowledgments}
This work was supported in part by the Spanish Ministry of Science and 
Technology (Grant FPA 2002-02037) and the University of the Basque Country 
(Grant 9/UPV00172.310-14497/2002).


\appendix

\section*{\label{sec:Coulomb}Appendix. Computation of $F_{\pm}(\bm{k})$}
\setcounter{section}{1}

In the following computation of $F_{\pm}(\bm{k})$, we shall
temporarily omit the indices $a,b$. 
We start with the evaluation of
$F_{+}(\bm{k})$ given by the contour integral of the Green's function
$G = (H-E)^{-1}$
\begin{equation}
  F_{+}(\bm{k})= \int_{C} \frac{dE}{2\pi i} e^{-\beta E} 
  \int d^2 \bm{r} e^{-i \bm{k}\cdot\bm{r}} 
  \langle\bm{r}|\frac{1}{H-E}|\bm{r}\rangle, 
\end{equation}
where the contour $C$ encircles clockwise the cut along the positive real $E$
axis and all the bound-state poles when $e^2<0$ in the case of an
attractive potential. 
To exploit the rotational symmetry of the problem and the properties
on momentum space, it proves convenient to average over the angle of $\bm{k}$
and compute the spatial integral making use of
\begin{equation}
 \int_{0}^\infty dr \, r J_{0}(k r) J_{0}(|\bm{p}-\bm{p}'|r) = 
 \frac{1}{k} \delta(k-|\bm{p}-\bm{p}'|). 
\end{equation}
This produces the expression
\begin{equation}
F_{+}(\bm{k})= \int_{C} \frac{dE}{2\pi i} e^{-\beta E} G_{+}(k,E),  
\end{equation}
where
\begin{equation}\label{eq:gplus}
    G_{+}(k,E)= \int d^2 \bm{p}\, d^2 \bm{p}'\,
\langle\bm{p}|\frac{1}{H-E}|\bm{p}'\rangle \frac{1}{2\pi k}
\delta(k-|\bm{p}-\bm{p}'|) .
\end{equation} 

The key to the computation of $G_{+}(k,E)$  is the closed form
formula for the momentum space Green's function
$G(\bm{p},\bm{p}')\equiv \langle\bm{p}|(H-E)^{-1}|\bm{p}'\rangle$
of the two-dimensional Coulomb problem.  This form has been given 
in \cite{Dittrich} following the clever treatment of Schwinger~\cite{Schwinger}.  
Let us summarize
the main results concerning this Green's function.  The momentum
representation equation for the Green's function
is
\begin{equation}
 \left(\frac{\bm{p}^2}{2m}-E \right)G(\bm{p},\bm{p}') +
 \frac{e^2}{2\pi}\int d^2 \bm{p}'' \frac{1}{|\bm{p}-\bm{p}''|} G(\bm{p}'',\bm{p}') = 
 \delta(\bm{p}-\bm{p}'). 
\end{equation} 
Based on the (conformal) correspondence between the points of the euclidean
two-dimensional momentum space and the surface of the unit
three-dimensional sphere
\begin{eqnarray}
    p_{x} &=& \frac{p_{0}}{2 \left(\cos \theta/2\right)^2} 
         \sin\theta \cos\phi, \\
    p_{y} &=& \frac{p_{0}}{2 \left(\cos \theta/2\right)^2}
         \sin\theta \sin\phi ,
\end{eqnarray}
the element of area on the sphere and the two-dimensional measure in the 
plane are related by  
\begin{equation}\label{eq:medida}
  d\Omega = \left(\frac{2 p_{0}}{p_{0}^2 + p^2} \right)^2 d^2 \bm{p}, 
\end{equation}
where $p_{0}=\sqrt{-2 m E}$ is the positive real wavenumber corresponding to the 
negative energy $E$.  The angular distance $\gamma$ between the orientations of
the two points $\Omega=(\theta, \phi)$, $\Omega'=(\theta', \phi')$ on
the unit sphere is related to the euclidean distance between the
corresponding points $\bm{p}$ and $\bm{p}'$ by
\begin{equation}
 \left(2 \sin\frac{\gamma}{2}\right)^2= \frac{4 p_{0}^2}{(p_{0}^2+p^2)(p_{0}^2+p'^2)}
 |\bm{p}-\bm{p}'|^2 .
\end{equation}
Then, noting that the term $|\bm{p}-\bm{p}''|^{-1}$ can be written as 
\begin{equation}
 \frac{1}{|\bm{p}-\bm{p}''|} = 2 p_{0}(p_{0}^2+p^2)^{-1/2} (p_{0}^2+p''^2)^{-1/2}
 \sum_{l,m}\frac{4 \pi}{2 l+1}Y_{l}^m(\Omega) Y_{l}^{m}(\Omega'')^\ast, 
\end{equation}
and the completeness relation of the spherical harmonics
\begin{equation}
    \delta(\Omega-\Omega') = 
    \sum_{l,m} Y_{l}^m(\Omega) Y_{l}^{m}(\Omega')^\ast , 
\end{equation}
the Green's function is just
\begin{equation}\label{eq:green}
  G(\bm{p},\bm{p}') = 8 m p_{0}^2\, (p_{0}^2+p^2)^{-3/2} (p_{0}^2+p'^2)^{-3/2}
  \sum_{l,m}\frac{Y_{l}^m(\Omega) Y_{l}^{m}(\Omega')^\ast}{1 + \frac{\nu}{l+1/2}} ,
\end{equation}
where 
\begin{equation}
\nu = \frac{e^2 m}{p_{0}}
\end{equation}
is positive when the potential is repulsive.  For an attractive
potential ($\nu<0$), the poles corresponding to the bound states are
\begin{equation}
   E_{l}= -\frac{m e^4}{2(l+1/2)^2} .
\end{equation}

The next step is to express the remaining factor in the integrand 
of (\ref{eq:gplus}) in terms of the spherical harmonics associated
with $\bm{p}$ and $\bm{p}'$.  This is readily performed from the
series of Legendre polynomials 
\begin{equation}
 \delta(2-2\cos\gamma-a^2)= \sum_{l=0}^\infty \frac{2l+1}{4} 
 P_{l}\left(1-\frac{a^2}{2}\right) P_{l}(\cos \gamma), 
\end{equation}
with the identification 
\begin{equation}
   a^2 = \frac{4 p_{0}^2 k^2}{(p_{0}^2+p^2)(p_{0}^2+p'^2)},   
\end{equation}
where $\gamma$ is the angle between the orientations in the unit
sphere corresponding to $\bm{p}$ and $\bm{p}'$.  Therefore, using the
addition of spherical harmonics
\begin{equation}\label{eq:add}
  P_{l}(\cos \gamma) = 
  \frac{4\pi}{2l+1}\sum_{m=-l}^l Y_{l}^m(\Omega) Y_{l}^{m}(\Omega')^\ast, 
\end{equation}
one obtains the required formula
\begin{eqnarray}
 \frac{1}{2 \pi k} \delta(k-|\bm{p}-\bm{p}'|)  &=& 
 \frac{4 p_{0}^2}{(p_{0}^2+p^2)(p_{0}^2+p'^2)} \sum_{l,m}
 P_{l}\left(1-\frac{a^2}{2}\right) 
 Y_{l}^m(\Omega) Y_{l}^{m}(\Omega')^\ast \nonumber \\
 &=& \left(\frac{2 p_{0}}{p_{0}^2+p^2}\right)^2 \delta(\Omega-\Omega') + 
 \mathrm{O}(k^2).  
\end{eqnarray}

With these results in hand, we return to the computation of $G_+(k,E)$. 
Using the integral representation of Legendre polynomials \cite{Abramo} 
\begin{equation}
P_l(\cos \gamma)=\frac{1}{\pi}\int_0^\pi d\phi\left(\cos\gamma+i \sin\gamma \cos\phi \right)^l, 
\end{equation}
and the hypergeometric series with $b=\nu+1/2$ and $z=\cos\gamma+i \sin\gamma \cos\phi$
\begin{equation}
\sum_{l=0}^\infty \frac{z^l}{l+b}= F(b,1,b+1,z), \quad (|z| < 1)
\end{equation}
the sum over $l$ in the Green's function~(\ref{eq:green}) 
can be easily written in an appropriate form  to study the $\bm{k}\rightarrow 0$ or, 
equivalently, the $\gamma \rightarrow 0$ behavior. 
This form is based on the following transformation 
formula 
for $F(b,1,b+1,z)$, valid when $|\arg(1-z)|<\pi$ and $|1-z|<1$
(see, for example, p 559 of \cite{Abramo})
\begin{equation}
\fl
F(b,1,b+1,z) = b \sum_{n=0}^\infty \frac{(b)_n}{n!}\left[
\psi(n+1)-\psi(b+1)-\ln(1-z)\right](1-z)^n .
\end{equation}
Therefore, by evaluating the $\phi$ integral of the $n=0$ term, we obtain
\begin{equation}
\sum_{l=0}^\infty \frac{P_l(\cos\gamma)}{l+\nu+1/2} = 
-\gamma_{\mathrm{E}} - \psi\left(\frac{1}{2}+\nu\right)-
\ln\left(\frac{\gamma}{2}\right) + O(\gamma \ln \gamma) , 
\end{equation}
where $\gamma_{\mathrm{E}}$ is the Euler's constant and $\psi$ denotes the digamma function. 
Also, we need
\begin{equation}
\sum_{l=0}^\infty P_l(\cos\gamma) =\frac{1}{2 \sin\gamma/2} .  
\end{equation}
Putting all pieces together, the non-vanishing portion of 
$G_+(k,E)$ as $\bm{k}\rightarrow0$ is readily found by the 
integration of (\ref{eq:gplus}). 
This yields, after subtraction of the $e^2=0$ contribution, the result
\begin{eqnarray}\label{gplus1}
\fl
G_+(k,E)=-\frac{2 m \nu}{p_0 k} - \frac{m \nu^2}{p_0^2}\left[
2\gamma_{\mathrm{E}}-1+2\ln\left(\frac{k}{p_0}\right) +
2 \psi\left(\frac{1}{2}+\nu\right) \right]+
O(k) \nonumber \\
\lo= \beta\left\{-\frac{2\pi y}{\lambda k (-t)} -\frac{\pi y^2}{2(-t)^2}
\left[2\gamma_{\mathrm{E}}-1+\ln\left(\frac{\lambda^2 k^2}{4\pi}\right) \right.\right. \nonumber \\
  \left.\left. - \ln(-t)
+ 2\psi\left(\frac{1}{2}+\frac{\sqrt{\pi} y}{\sqrt{-t}}\right)\right]+O(k)\right\}, 
\end{eqnarray} 
where $t = \beta E$, and the dimensionless parameter $y$ is the ratio of the Coulomb energy for 
two particles separated by one thermal (reduced) wavelength to the temperature
\begin{equation}
y = 
\pm \sqrt{\frac{\beta m e^4}{2\pi}} \Leftrightarrow 
\eta_{ab}= \frac{\beta e_a e_b}{\lambda_{a b}}.  
\end{equation}
Thus, the $\pm$ sign accounts for the repulsive ($+$) or attractive ($-$) cases. 
Note that $\arg(-t)$ is purely real when $t$ is on the negative part of the real $t$ axis, 
so that, the convention is $-\pi < \arg(-t) < \pi$ when 
$0 < \arg(t) < 2\pi$. 

It remains to perform the required contour integral. 
The result $F_+(k)$ parametrically depends on $y$, and we will distinguish between the 
small and large $|y|$ limits.
For both the attractive and repulsive cases, the small $|y|<<1$ limit can be obtained 
from the power series expansion of the $\psi$ function
\begin{equation}
\fl
\psi(1/2 + z)=-\gamma_{\mathrm{E}} - 2 \ln2 + \sum_{n=2}^\infty \zeta(n)(-1)^n (2^n-1) z^{n-1}, 
\quad \left(|z|<\frac{1}{2}\right)
\end{equation}
provided that the contour $C$ cuts the real $t$ axis at the left of $t=-4\pi y^2$, 
in the case of an attractive potential, or at the left of $t=0$, 
in the case of a repulsive potential. 
The insertion term by term of this series in the Hankel's expression for $\Gamma(z)$  
\begin{equation}
\frac{1}{\Gamma(z)} = \int_C \frac{dt}{2\pi i} (-t)^{-z} e^{-t} , 
\end{equation} 
and its derivative with respect to $z$
\begin{equation}
\frac{\psi(z)}{\Gamma(z)} = \int_C \frac{dt}{2\pi i} \ln(-t) (-t)^{-z} e^{-t} , 
\end{equation} 
produces the final result, after subtraction of the $e^2=0$ contribution 
\begin{equation}\label{eq:fplus}
    \Delta F_+(\bm{k}) = -\frac{2\pi y}{\lambda k} - 
    \frac{\pi y^2}{2}\left[ \gamma_{\mathrm{E}} +
\ln\left(\frac{\lambda^2 k^2}{64\pi} \right) + f(y)\right] 
+ O(k) ,
\end{equation}
where the function $f(y)$ is 
defined by the power series expansion
\begin{equation}
\fl
f(y) = -2 -  
2\sum_{n=2}^\infty 
\frac{\zeta(n)(2^n-1)}{\Gamma((3+n)/2)}
\left(-\sqrt{\pi} y \right)^{n-1}= -2+\frac{4\pi^2 y}{3}-7\pi \zeta(3) y^2
+ O(y)^3.
\end{equation} 
Note that here $y$ can be positive or negative. 

Now, consider the asymptotic behavior when $|y|\rightarrow \infty$. 
For the attractive case, the leading contribution comes from the lowest bound state, 
as it is easily noted from the dependence on $e^{-\beta H}$. 
Hence, all we need is the residue $-e^{4 \pi y^2}$ of the 
integrand $e^{-t} G_+(k,E)$  at the bound state pole $t=-4 \pi y^2$. Thus, 
when $y\rightarrow -\infty$,
\begin{equation}
\Delta F_+(\bm{k})  \sim e^{4 \pi y^2}, 
\end{equation}
and 
\begin{equation}
 f(y) \sim -\frac{2}{\pi y^2} e^{4 \pi y^2}. 
\end{equation}

To obtain the leading asymptotic behavior in the case of strong repulsion, 
it proves convenient to study 
separately the contribution counterclockwise around the ramification point $t=0$, 
and the contribution from the discontinuity of $G_+(k,E)$ at the cut along the positive real axis. 
As we shall see, the discontinuity yields exponentially small corrections. 
Using the identity
\begin{equation}
\mathrm{Im}\,\psi(1/2 + i u) = \frac{\pi}{2} \tanh(\pi u), 
\end{equation} 
and noting that $\mathrm{Disc}\,\ln(-t)= -2\pi i$, the discontinuity of $G_+(k,E)$ becomes
\begin{eqnarray}\label{eq:disco}
\fl
G_+(k,E+i \varepsilon)-G_+(k,E-i \varepsilon) &=& \frac{i \pi^2 y^2}{t^2} 
\left[-1-\tanh\left(\frac{\pi^{3/2} y}{\sqrt{t}} \right) \right] \nonumber \\
 &=& -\frac{2 i \pi^2 y^2}{t^2} + \frac{i \pi^2 y^2}{t^2} 
\left[1-\tanh\left(\frac{\pi^{3/2} y}{\sqrt{t}} \right) \right], \; \left(t>0 \right). 
\end{eqnarray}
The last term in the second line produces an integral well defined for $y>0$, 
which can be studied by inserting the 
representation of $\tanh z$ as a series of exponentials. The resulting expression contributing to 
$\Delta F_+(\bm{k})$ has the form 
\begin{eqnarray}
&&-\pi y^2 \sum_{n=0}^\infty (-1)^n \int_0^\infty \frac{dt}{t^2} 
\exp\left(-t - \frac{2(n+1) \pi^{3/2} y}{\sqrt{t}}\right)  \nonumber \\
&&\qquad \sim-\frac{2y}{\sqrt{3}}e^{-3 \pi y^{2/3}} \left(1 + \frac{17}{36 \pi y^{2/3}}+
\ldots \right) ,  
\end{eqnarray}
where, as $y \rightarrow \infty$, the leading behavior corresponds to the first term in the sum, 
which has been  evaluated by the steepest descent approximation. 
The first term in (\ref{eq:disco}) yields
\begin{equation}\label{eq:dive}
-\pi y^2 \int_{\delta}^\infty  \frac{dt}{2\pi} \frac{e^{-t}}{t^2} = 
-\frac{\pi y^2}{\delta}  + \pi y^2 (1-\gamma_{\mathrm{E}}-\ln\delta) + O(\delta). 
\end{equation} 
It remains to compute the contribution from the small circle around the origin. 
There, $-t= \delta e^{i \theta}$, with $-\pi<\theta<\pi$, and $\ln(-t) = \ln\delta + i\theta$. 
Then, the contribution to  $\Delta F_+(\bm{k})$ may be obtained  by 
using the large $z$ behavior of the $\psi$ function
\begin{equation}
\psi(1/2 + z) \sim \ln z + \sum_{n=1}^\infty \frac{B_{2 n}}{2 n}\left(1-2^{1-2n}\right) z^{-2n},
\quad (|\arg(z)|<\pi )
\end{equation}
where $B_{2 n}$ are the Bernoulli numbers. The angular integral of $e^{-t} G_+(k,E)$ 
produces singular terms which serve to cancel the linear and logarithmic divergences 
in (\ref{eq:dive}), in addition to non-vanishing terms when $\delta\rightarrow 0$. 
Putting the pieces together, one immediately finds 
all terms of  $\Delta F_+(\bm{k})$ non-exponentially suppressed in the large $y$ limit, 
\begin{equation}
\Delta F_+(\bm{k})\sim -\frac{\pi y^2}{2}\left[4 \gamma_{\mathrm{E}} -3 + 
 \ln\left(\frac{\lambda^2 k^2 y^2}{4} \right) \right] - 
\frac{2\pi y}{\lambda k} -\frac{1}{24}. 
\end{equation}
Note that the only contribution of the asymptotic expansion of the $\psi(1/2+z)$ function 
comes from the logarithmic and the $z^{-2}$ terms.  With this result,
the asymptotic behavior of $f(y)$ as $y\rightarrow \infty$ is given by
\begin{equation}\label{eq:strong}
f(y) \sim \ln\left(16\pi y^2 \right) + 3 \gamma_{\mathrm{E}}-3 +
\frac{1}{12\pi y^2}. 
\end{equation}

It remains to compute the exchange contribution. 
The nonvanishing portion at $\bm{k}=0$  of the exchange contribution is given by 
\begin{equation}
F_{-}(\mathbf{0})= \int_{C} \frac{dE}{2\pi i} e^{-\beta E} G_{-}(0,E),  
\end{equation}
where
\begin{equation}\label{eq:gminus}
    G_{-}(0,E)= \int d^2 \bm{p}\, G(\bm{p},-\bm{p}) .
\end{equation} 
Noting that the  orientations on the unit sphere of the points $\bm{p}$ and $-\bm{p}$ 
are $\Omega=(\theta,\phi)$ and 
$\Omega'=(\theta,\phi+\pi)$ respectively, the angular distance between them is $2\theta$. 
Thus, using (\ref{eq:add}), 
the sum over $m$ in (\ref{eq:green}) gives a factor $P_l(\cos 2\theta)$. 
Now, the integration over $\bm{p}$ can be performed with the aid of the following 
integrals of Legendre polynomials  
\begin{equation}
\int_0^\pi d\theta \sin\theta P_l(\cos 2\theta)\cos^2 \theta/2 = \frac{(-1)^l}{2l+1},
\end{equation}
and, finally, the resulting series is summed using the series representation
\begin{equation}
\sum_{l=0}^\infty \frac{(-1)^l}{2 l + 1 + 2 \nu}=
\frac{1}{4}\left[\psi\left(\frac{3}{4}+\frac{\nu}{2}\right) - 
                 \psi\left(\frac{1}{4}+\frac{\nu}{2}\right)\right] .
\end{equation}
Hence, subtracting  the $e^2=0$ contribution, $G_{-}(\mathbf{0},E)$ reads
\begin{equation}
G_{-}(0,E) = \frac{m \nu}{2 p_0^2}\left[
                 \psi\left(\frac{1}{4}+\frac{\nu}{2}\right) - 
                 \psi\left(\frac{3}{4}+\frac{\nu}{2}\right) \right] .
\end{equation}
 
Since the exchange contribution involves particles of the same species, 
only the computation of $F_{-}(\mathbf{0})$  
in the repulsive case $y>0$ is required. 
In this case, we can use the integral representation
\begin{equation}\label{eq:cosh}
\psi\left(\frac{3}{4}+z\right) - \psi\left(\frac{1}{4}+z\right) = 
\frac{1}{2} \int_0^\infty du \frac{e^{-z u}}{\cosh u/4}\,. \quad  (\mathrm{Re}\, z > 0)
\end{equation}
When $y\ll 1$, we insert the power series expansion of the exponential $e^{-z u}$ and 
perform the $u$ integral with the aid of 
\begin{equation}
\int_0^\infty du \frac{u^n}{\cosh u/4} = 
2 \Gamma(n+1) [\zeta(n+1,1/4) -\zeta(n+1,3/4)], \quad  (\mathrm{Re}\, n > -1)
\end{equation} 
where $\zeta(s,a)$ denotes the generalized Riemann zeta function. 
Then, the evaluation of the contour integral yields the power series expansion 
\begin{equation}
\fl
    F_{-}(\mathbf{0}) = -\frac{\pi y}{2} -
\sum_{n=1}^\infty
\frac{(-1)^n[\zeta(1+n,1/4)-\zeta(1+n,3/4)]}{2\Gamma\left((3+n)/2\right)}
\left(\frac{\sqrt{\pi}\, y}{2}\right)^{n+1} .
\end{equation}
Similarly to the function $f$ for the direct part, it is useful to
define the function $\widetilde{f}(y)$ as
\begin{equation}\label{eq:fminus}
 \widetilde{f}(y) \equiv -\frac{2}{\pi y^2} F_{-}(\mathbf{0})=
 \frac{1}{y} - 4 C +\frac{\pi^3 y}{3}+ O(y^2), 
\end{equation}
where $C$ is the Catalan's number. 

Finally, the positive large $y$ limit 
may be obtained  
by evaluating the $u$ integral~(\ref{eq:cosh}) using the 
Laplace's method of asymptotic analysis. Inserting the Taylor expansion 
of $(\cosh u/4)^{-1}$ about $u=0$, produces a convergent series 
in $u < 2 \pi$ which can be integrated term-by-term. 
Therefore, the Hankel's formula leads to 
\begin{equation}
F_{-}(\mathbf{0}) \sim -\frac{1}{4}, 
\end{equation}
and 
\begin{equation}\label{eq:stroex}
\widetilde{f}(y) \sim \frac{1}{2 \pi y^2},
\end{equation}
apart from exponentially small corrections.


\end{document}